\documentclass[aip,jcp,floatfix]{revtex4-1}
\usepackage{placeins}
\usepackage[utf8]{inputenc}
\usepackage{multirow}
\usepackage{array}
\usepackage{booktabs}
\usepackage{tabularx}
\usepackage{makecell}
\usepackage{graphicx}
\usepackage{xcolor}
\usepackage{enumitem}
\usepackage{dcolumn}
\usepackage{amsmath}
\usepackage{physics}
\newcolumntype{C}[1]{>{\centering\arraybackslash}p{#1}}
\def\mic{CH$_{2}$NH$_{2}^{+}$}
\newcommand{\ppci}{$\pi\pi^*/S_0$}
\newcommand{\spci}{$\sigma\pi^*/S_0$}
\newcommand{\ssci}{$S_2/S_1$}
\usepackage{float}
\usepackage{siunitx} 

\draft 

\begin{document}


\title{XMCQDPT2-Fidelity Transfer-Learning Potentials and a Wavepacket Oscillation Model with Power-Law Decay for Ultrafast Photodynamics}



\author{Ivan V. Dudakov}
\affiliation{Department of Chemistry, Lomonosov Moscow State University, Leninskie Gory 1/3, Moscow 119991, Russia}
\affiliation{MSU Institute for Artificial Intelligence, Lomonosov Moscow State University, Moscow 119192, Russia}

\author{Pavel M. Radzikovitsky}
\affiliation{Department of Chemistry, Lomonosov Moscow State University, Leninskie Gory 1/3, Moscow 119991, Russia}

\author{Dmitry S. Popov}
\affiliation{Department of Chemistry, Lomonosov Moscow State University, Leninskie Gory 1/3, Moscow 119991, Russia}

\author{Denis A. Firsov}
\affiliation{Department of Chemistry, Lomonosov Moscow State University, Leninskie Gory 1/3, Moscow 119991, Russia}

\author{Vadim V. Korolev}
\affiliation{Department of Chemistry, Lomonosov Moscow State University, Leninskie Gory 1/3, Moscow 119991, Russia}
\affiliation{MSU Institute for Artificial Intelligence, Lomonosov Moscow State University, Moscow 119192, Russia}

\author{Daniil N. Chistikov}
\affiliation{Department of Chemistry, Lomonosov Moscow State University, Leninskie Gory 1/3, Moscow 119991, Russia}
\affiliation{Institute of Quantum Physics, Irkutsk National Research Technical University, 83 Lermontov Street, Irkutsk 664074, Russia}

\author{Vladimir E. Bochenkov}
\affiliation{Department of Chemistry, Lomonosov Moscow State University, Leninskie Gory 1/3, Moscow 119991, Russia}

\author{Anastasia V. Bochenkova}
\email{bochenkova@phys.chem.msu.ru}
\affiliation{Department of Chemistry, Lomonosov Moscow State University, Leninskie Gory 1/3, Moscow 119991, Russia}


\date{\today}

\begin{abstract}
A central pursuit in theoretical chemistry is the accurate simulation of photochemical reactions, which are governed by nonadiabatic transitions through conical intersections. Machine learning has emerged as a transformative tool for constructing the necessary potential energy surfaces, but applying it to excited states faces a fundamental barrier: the prohibitive cost of generating high-level quantum chemistry data. We overcome this challenge by developing machine-learning interatomic potentials (MLIPs) that achieve multi-state multi-reference perturbation theory accuracy through various techniques, such as transfer, multi-state, and $\Delta$-learning. Applied to the methaniminium cation, a fundamental model system in photochemistry, our highest-fidelity transfer-learning model uncovers its complete photodissociation landscape following S$_2$ photoexcitation. The comprehensive XMCQDPT2/SA(3)-CASSCF(12,12) electronic structure description captures all competing decay channels, including S$_1$ branching into photoisomerization and direct H$_2$-loss pathways. Our results show that the population dynamics generally depends on the MLIP model, correlating with its performance. At the same time, the introduction of MLIP-uncertainty corrections based on the predictions of an ensemble of models brings different approaches into agreement, validating this metric as essential for reliable dynamics. To interpret the population dynamics, we introduce a wavepacket oscillation model -- a mechanistically transparent, power-law kinetics framework that extracts state-specific lifetimes directly from first-principles simulations. The model quantitatively reproduces the ultrafast decay in \mic{}, creating a direct link between quantum transition probabilities and classical rate constants. The kinetic fits yield channel-specific lifetimes, supporting the recently discovered photochemical pathway mediated by a novel \spci{} conical intersection.
\end{abstract}

\pacs{}

\maketitle 

\section{Introduction}
Reliability, convergence, and physical soundness of machine learning (ML)-driven nonadiabatic molecular dynamics (NAMD) simulations depend heavily on the accuracy of the underlying ML interatomic potential (MLIP) in approximating the potential energy surfaces of both ground and excited states. As of today, neural networks (NNs) are the primary MLIP algorithms, superseding the alternative formulations (\textit{e.g.}, kernel methods)~\cite{Muller2025}. Architecturally, most advances in excited-state NN-MLIPs have been adapted from ground-state modeling. Among these, models using atom-wise descriptors, including invariant types (\textit{e.g.}, Behler–Parrinello NNs~\cite{behler2007generalized}, ACE\cite{drautz2019atomic}, and ANI~\cite{devereux2020extending}) and equivariant types (\textit{e.g.}, PaiNN~\cite{schutt2021equivariant}, MACE\cite{batatia2022mace}, and NequIP~\cite{batzner20223}), currently deliver state-of-the-art performance across diverse chemical systems.
However, simulating excited states introduces unique challenges not present in ground-state systems. These include accurately capturing nonadiabatic couplings, describing multiple interacting electronic states, and properly modeling conical intersections that govern photochemical pathways. Furthermore, unlike ground-state MLIPs typically trained on density functional theory (DFT) data, advanced models for studying photoinduced phenomena must approximate a set of potential energy surfaces from computationally demanding wavefunction-based methods.~\cite{Westermayr2022,Fang2024} 
Consequently, diverse sampling techniques have been developed to construct compact yet representative datasets, ensuring MLIPs remain in an interpolative regime during long-time-scale simulations.\cite{westermayr2019machine,li2021automatic} These approaches typically begin with key stationary points, including minima and transition states, along with Wigner-sampled conformations and NAMD trajectories. This initial sampling is then refined through iterative active learning~\cite{Bi2025} to systematically improve coverage of underrepresented critical regions. To ensure robustness, active learning typically targets geometries with high predictive uncertainty for additional reference calculations, though domain-specific selection criteria are gaining traction.\cite{Schwalbe-Koda2021} Beyond the challenge of conformational sampling, the need to approximate multiple potential energy surfaces has driven the development of multi-state models. These architectures explicitly capture complex interstate correlations, leading to distinct improvements in the accuracy and transferability of excited-state MLIPs.\cite{martyka2025charting,barrett2025transferable} 

These methodological advances are primarily driven by the challenge of constructing excited-state potential energy surfaces under constrained computational resources. Crucially, an MLIP accuracy is inherently bounded by its training data; even an optimal model reproduces and thus inherits the limitations of the underlying quantum chemistry method. Developing robust MLIPs for NAMD requires data that accurately captures both static and dynamic electron correlation across all geometries. This presents a challenge: single-reference post-Hartree-Fock methods fail for molecular configurations far from equilibrium, while multi-configuration methods like CASSCF lack dynamic correlation, limiting the applicability and accuracy of both standard approaches for training excited-state MLIPs. 

Furthermore, direct application of pioneering models for high-fidelity studies is not always straightforward. The universal OMNI-P2x potential\cite{martyka2025omni}, for instance, is built upon single-reference TD-DFT data, which fundamentally limits its ability to accurately describe the multi-configurational character of potential energy surfaces near conical intersections. Likewise, the advanced X-MACE model\cite{barrett2025transferable}, while highly accurate, is inherently tied to the fidelity of its specific training dataset and is tailored for dynamics protocols requiring properties like non-adiabatic couplings. This creates a persistent need for developing custom-built potentials when the scientific goal is to achieve the accuracy of a particular high-level quantum chemistry benchmark or to employ alternative simulation protocols that do not rely on such auxiliary properties.

In the present work, we develop a series of MLIPs that achieve the accuracy of multi-state multi-reference perturbation theory, specifically targeting the extended multi-configuration quasi-degenerate perturbation theory (XMCQDPT2~\cite{xmcqdpt2}) level. Renowned for its reliability, the XMCQDPT2 method correctly describes the topology and topography around conical intersections and excited-state energy barriers along photochemical pathways.~\cite{BOCHENKOVA2024141} This is evidenced by its success in predicting statistical barrier-controlled excited-state lifetimes for key biological chromophores, including the retinal protonated Schiff-base from rhodopsins~\cite{Kiefer2019NC,Gruber2021} and the green fluorescent protein chromophore~\cite{GFP_JACS}. By implementing a suite of advanced techniques, including transfer learning,\cite{smith2019approaching} multistate learning,\cite{martyka2025charting} and $\Delta$-learning,\cite{ramakrishnan2015big} we significantly boost predictive performance of MLIPs, thereby improving the reliability of ML-driven NAMD simulations. 

We employ the developed framework to map the competing excited-state decay channels in the nonadiabatic dynamics of the methaniminium cation (\mic{}), a fundamental model system in photochemistry~\cite{barbatti2007fly,tapavicza2007trajectory,yamazaki2005locating,suchan2020pragmatic,pittner2009optimization,west2014nonadiabatic, barbatti2006ultrafast, fabiano2008approximate, westermayr2019machine, westermayr2020combining,lan_jade}. As the simplest protonated Schiff base, this cation serves as a model for biological chromophores like retinal. Furthermore, it is a key atmospheric precursor on Titan, where it leads to the formation of HCNH$^+$, the most abundant ion in its upper atmosphere.~\cite{Nixon2024,singh2010photodissociation,pei2012ion,thackston2018quantum} Building on our initial discovery of a UV-driven direct H$_2$ elimination pathway in \mic{} via a novel \spci{} conical intersection\cite{Bochenkova_JPCL2025}, we now present a refined dynamical picture of its photoinduced dynamics using XMCQDPT2-fidelity machine learning potentials. Our large-scale simulations, enhanced with out-of-sample diagnostics, provide the time evolution of state populations and the branching ratios of photodissociation products for dynamics initiated from the optically bright S$_2$ state, which is of valence $\pi\pi^*$ character in \mic{}. This work presents, to our knowledge, the first on-the-fly nonadiabatic dynamics simulations for this system at a high level of theory, employing a large (12e,12o) active space that encompasses all valence $\sigma$ and $\pi$ orbitals. This comprehensive electronic structure description is essential to capture all competing photodissociation channels simultaneously and accurately. Finally, to move beyond empirical fitting and directly connect the microscopic nonadiabatic dynamics to ensemble kinetics, we introduce a multi-passage survival cascade model. This model quantifies nonadiabatic decay through the probability of surviving multiple passages through a conical intersection, providing a mechanistically transparent, power-law framework to extract state-specific lifetimes from first-principles population dynamics. 

\section{Methods}
\subsection{Reference electronic-structure data}
Training data for implementing machine learning interatomic potentials (MLIPs) was obtained at the XMCQDPT2/SA(3)-CASSCF(12,12)/aug-cc-pVDZ level of theory using the Firefly quantum chemistry package\cite{Firefly}. Since the computational cost of XMCQDPT2 calculations precludes exhaustive trajectory sampling, a compact yet diverse training set was constructed as follows. First, we selected a representative subset of 1,000 geometries from our previous SA(3)-CASSCF(12,12)/aug-cc-pVDZ dataset containing $\sim$50,000 configurations\cite{Bochenkova_JPCL2025}. Using a greedy maximin (farthest-point) sampling approach in Cartesian coordinate space, we maximized the minimum pairwise distance between selected points to ensure broad coverage of the conformational landscape explored during our earlier on-the-fly NAMD simulations. For each selected geometry, XMCQDPT2 single-point energies and energy gradients were computed for the lowest three CH$_2$NH$_2^+$  singlet electronic states, herein denoted S$_0$, S$_1$, and S$_2$. Second, to improve the representation of potential energy surfaces near conical intersections, we augmented the dataset with 450 additional configurations, specifically, 150 samples from the vicinity of each of the three conical intersections. Starting from the XMCQDPT2-optimized MECI structures, we generated these points by sampling small displacements within the branching plane, which is defined by the gradient difference and nonadiabatic coupling vectors.

Because temporally correlated NAMD configurations can lead to overoptimistic generalization estimates (assuming random training-validation-test split), we enforced trajectory-wise splits. For the 1,000 XMCQDPT2 points inherited from the CASSCF pool, we preserved the original grouping and subset assignments, thereby mirroring the CASSCF distribution in the conformational space. The MECI-proximal XMCQDPT2 configurations were split randomly across subsets because they do not originate from trajectories. This procedure prevented trajectory-level leakage into the test set and enabled more faithful out-of-trajectory generalization assessment.

\subsection{Neural network potentials}
We employed the Multi Atomic Cluster Expansion (MACE) architecture \cite{batatia2022mace} as a basic model for approximating three lowest-energy singlet potential energy surfaces. MACE is an equivariant message-passing neural network that assigns atomic energy contributions $\epsilon_i$ from local environments and obtains the total state energy as $E = \sum_i \epsilon_i$. Energy gradients are computed by automatic differentiation of $E$ with respect to atomic coordinates.

\subsection{Training protocols}
We investigated four complementary modeling approaches, varying both the treatment of interstate coupling within the MLIP architecture and the strategy for initializing the model weights.

\begin{enumerate}[leftmargin=*]
    \item \textbf{Single-state models with random initialization (SS-RI).} Three separate MACE models were trained independently for S$_0$, S$_1$, and S$_2$ states. Each model took Cartesian coordinates as input and predicted the corresponding state energy and gradient. Model weights were initialized randomly.
    \item \textbf{Single-state transfer learning from CASSCF to XMCQDPT2 (SS-TL).} We first trained state-specific MACE models on the large CASSCF datasets using the same architecture and loss function. These CASSCF-trained weights were then used to initialize the corresponding single-state models for XMCQDPT2, which were subsequently finetuned on the XMCQDPT2 data. This transfer-learning protocol leverages representation reuse across related electronic-structure levels and is well aligned with recent practice in quantum chemical $\Delta$- and transfer learning\cite{ramakrishnan2015big, smith2019approaching, li2024construction}. Relative to random initialization, this strategy produced consistently improved test-set metrics.
    \item \textbf{Single-state $\Delta$-learning (SS-$\Delta$).} For each state, we trained a MACE model to predict the difference between target and baseline values, i.e., $\Delta E = E_{XMCQDPT2}-E_{CASSCF}$ and $\Delta F = F_{XMCQDPT2}-F_{CASSCF}$. The same composite energy–gradient loss function was applied to the $\Delta$-learning targets\cite{Huang_2025}. This approach yielded improved accuracy for the excited states. However, practical NAMD simulations with $\Delta$-models entails evaluating the baseline CASSCF energy and gradient at every NAMD step and adding the learned correction, which partially offsets the computational gains.
    \item \textbf{Multi-state learning.} We further trained a single shared MACE encoder with three state-specific neurons in the final layer (one per singlet state), jointly minimizing the sum of state-wise energy–gradient losses\cite{Westermayr_2020, D4SC04164J}. Two initialization strategies were utilized: random (MS-RI) and transfer from a multistate CASSCF-trained model (MS-TL). We also explored a $\Delta$-learning variant of the multi-state model by assigning three heads to predict $\Delta E$ and $\Delta F$ with respect to CASSCF (MS-$\Delta$).
\end{enumerate}

\subsection{Training details}
The pretraining on CASSCF data was carried out with the following hyperparameters: a maximum spherical harmonics order of 3, 128 embedding channels, and a radial cutoff of 5.0~\AA. Training was performed for 1125 epochs using stochastic weight averaging \cite{izmailov2018averaging}. The loss function coefficients for energy and forces were set to 1:100 for the first 1125 epochs, and then changed to 1000:100 for the next 375 epochs to improve the accuracy of energy prediction after learning the energy gradient. The learning rates for these two stages were 10$^{-2}$ and 10$^{-3}$, respectively. The batch size was set to 32.

The models for XMCQDPT2 were trained with the same hyperparameters, but with a smaller number of epochs: 500 for single‑state models and 250 for multi-state models. The second training stage, with changed loss function coefficients, was started from the 375th and 186th epochs, respectively. For the finetuning stage, we used the same atomic reference energies as for the pretraining stage. Detailed information on predictive performance is provided in the Supplementary Material (Figs.~S1-S3).

\subsection{Nonadiabatic dynamics simulations}
For each training protocol, an ensemble of five neural network models was trained using identical datasets and hyperparameters. These models were tested by localization of three conical intersections, starting from the corresponding MECI structures located at the XMCQDPT2/SA(3)-CASSCF(12,12) level of theory. We then conducted two separate nonadiabatic dynamics simulations. The first employed an ensemble of models, which provided reasonable estimates of the MECI geometries. The second simulation used only the best-performing model, which was selected based on the smallest root-mean-square deviation between its predicted MECI structure and the XMCQDPT2 reference.

In order to simulate nonadiabatic molecular dynamics of the photoexcited methaniminium cation, we employed Landau-Zener surface hopping algorithm in a form proposed by Belyaev and Lebedev (LZBL) \cite{PhysRevA.84.014701}. An advantage of the LZBL method is that it does not require calculations of nonadiabatic couplings for the dynamics simulations. This feature enables an efficient machine-learning acceleration of the dynamics, as it would otherwise require fitting the surfaces for nonadiabatic couplings, which remains a major challenge. \cite{westermayr2019machine,westermayr2020combining,varga2018direct,li2021new,richardson2023machine}. At the same time, the LZBL method is capable of producing the results similar to those obtained by other methods\cite{suchan2020pragmatic}. The LZBL dynamics were performed using the MLatom package\cite{MLatom_paper}, which features an interface for dynamics simulations with MACE~\cite{batatia2022mace} neural network potentials. For each MLIP model, 600 trajectories were propagated with a 0.2~fs time step for a total duration of 100~fs per trajectory. initial conditions were sampled from the Wigner distribution around the S$_0$ equilibrium geometry. Ground-state optimization and vibrational analysis were conducted at the SA(3)-CASSCF(12,12) level of theory.

\subsection{Population correction via uncertainty quantification}
The robustness of the neural network potentials in sampling configurational space was assessed via out-of-sample diagnostics during ML/LZBL simulations. Predictive uncertainty was quantified using the standard deviation of energy predictions from a 5-model ensemble. To enhance the reliability of population predictions, we explicitly incorporated this uncertainty in our weighting algorithm. For each trajectory, we constructed a discrete indicator function for each electronic state. The value of the function was 1 if the system was in that particular state at a given time step, and 0 otherwise. The average population of a state was then the mean value of its corresponding indicator function over the ensemble of trajectories: 
\begin{equation}
\label{eq:pop_std_weighting}
p_{S_j}(t) = \frac{\sum\limits_{i=1}^Nw_i(t)\cdot\mathbf{1}_{S_j}(t)}{\sum\limits_{j=1}^K\sum\limits_{i=1}^Nw_i(t)\cdot\mathbf{1}_{S_j}(t)},
\end{equation}
where $p_{S_j}(t)$ is the average population of state $S_j$ at time $t$, $\mathbf{1}_{S_j}$ is the indicator function for state $S_j$, $w_i$ is the weight of trajectory $i$, $N$ is the total number of trajectories, and $K$ is the number of electronic states. To prioritize trajectories with more reliable energy predictions, we assigned each a weight inversely proportional to the standard deviation of the neural network ensemble. When all trajectory weights are set to unity, Eq.~\ref{eq:pop_std_weighting} reduces to the standard average, where the population of a given electronic state is simply the fraction of trajectories propagating on that state.

\subsection{Analysis of photodissotiation channels}
According to our previously developed protocol~\cite{Bochenkova_JPCL2025}, the trajectories originating from the S$_2$ state were classified into seven distinct groups:
\begin{enumerate}[leftmargin=*,topsep=0pt,itemsep=-1ex,partopsep=1ex,parsep=1ex]
    \item Cleavage of $\mathrm{CN}$ bond resulting in formation of $\mathrm{CH_2^+}$ and $\mathrm{NH_2}$ fragments; 
    \item Elimination of two hydrogen atoms from the carbon center, resulting in formation of $\mathrm{CNH_2^+}$;
    \item Elimination of two hydrogen atoms from the nitrogen center, resulting in formation of $\mathrm{CH_2N^+}$;
    \item Elimination of the hydrogen atom from the carbon center, resulting in formation of $\mathrm{HCNH_2^+}$;
    \item Elimination of the hydrogen atom from the nitrogen center, resulting in formation of $\mathrm{H_2CNH^+}$;
    \item Elimination of one hydrogen atom from both carbon and nitrogen centers, resulting in formation of $\mathrm{HCNH^+}$;
    \item disostiation of $\mathrm{H_2CNH_2^+}$ does not happen at the simulation timescale.\\
\end{enumerate}

We assigned trajectories to one of these groups according to the following procedure:
\begin{enumerate}[leftmargin=*,topsep=0pt,itemsep=-1ex,partopsep=1ex,parsep=1ex]
    \item For each trajectory, the $\mathrm{CN}$ bond length $d_\mathrm{CN}$ was measured. If the maximum value exceeded 3.55~{\AA}, trajectory was assigned to group 1 and excluded from subsequent analysis
    \item For each remaining trajectory, the maximum value of $f_\mathrm{CH_2}$ was calculated:
    \[   
    f_{\text{CH}_2} = \left[\frac{1}{2}\cdot\left(\frac{d_{\text{CH}^1}}{d_{\text{CH}^2}} + \frac{d_{\text{CH}^2}}{d_{\text{CH}^1}} \right)  \right]^{-1}\cdot d_{\text{H}^{12}_{cm}\text{C}},
    \]
    
     where $d_{\text{CH}^i}$ is the distance from the C atom to one of the H atoms which are bonded to the C atom at the equilibrium geometry; $d_{\text{H}^{12}_{cm}}$ is the distance from the center of mass of these hydrogen atoms to the C atom. The trajectories, for which $\max f_{\text{CH}_2}$ exceeded 1.95~{\AA}, were placed in the second group and excluded from subsequent analysis. 
    \item For each remaining trajectory, the maximum value of $f_{\text{NH}_2}$ was calculated: 
     \[
    f_{\text{NH}_2} = \left[\frac{1}{2}\cdot\left(\frac{d_{\text{NH}^3}}{d_{\text{NH}^4}} + \frac{d_{\text{NH}^4}}{d_{\text{NH}^3}} \right)  \right]^{-1}\cdot d_{\text{H}^{34}_{cm}\text{N}}
    \]
    where $d_{\text{NH}^i}$ is the distance from the N atom to one of the hydrogen atoms which are bonded to the N atom at the equilibrium geometry; $d_{\text{H}^{34}_{cm}}$ is the distance from the center of mass of these hydrogen atoms to the N atom. The trajectories, for which $\max f_{\text{NH}_2}$ exceeded 1.64~{\AA}, were placed in group 3 and excluded from subsequent analysis.
    \item 
    For remaining trajectories, the values of the two distances $d_{\text{CH}^1}$ and $d_{\text{CH}^2}$ were computed at every time step. If the maximum value of one of these distances exceeded 2.51~{\AA}, the trajectory was placed in group 4, except for those trajectories that also satisfied the conditions for being placed in group 5. 
    \item Similarly, if for any of the trajectories remaining after steps 1-3 the maximum value of one of the two distances $d_{\text{NH}^3}$ and $d_{\text{NH}^4}$ exceeded 1.95~{\AA}, the trajectory was placed in group 5, except for those trajectories that also satisfied the conditions for being placed in group 4. 
    \item If the trajectory simultaneously satisfied the conditions of groups 4 and 5, it was assigned to group 6.
    \item All remaining trajectories were assigned to group 7. 
\end{enumerate}
The cutoff values were established based on the temporal variation of the described parameters throughout the trajectories (see Supplementary Material, Fig.~S4).

\section{Results and Discussion}
\subsection{Model performance metrics}

Table~\ref{tab:metrics} provides an evaluation of how different training strategies influence the accuracy of energy and energy gradient predictions on the XMCQDPT2 test set for \mic{}. Transfer learning yields the most substantial improvement, reducing the mean absolute error (MAE) for energies (and energy gradients) by 28.6\% (29.5\%), 20.4\% (25.2\%), and 21.3\% (20.1\%) compared to models trained from scratch with random weight initialization. This enhancement is achieved through pretraining on a diverse dataset containing CASSCF data. While multi-state learning and $\Delta$-learning also reduced error metrics overall, we observed unexpected performance degradation for specific quantities. Notably, the S$_0$-state energy MAE increased by 50-150\% for the MS-RI, MS-TL, and MS-$\Delta$ models, and the S$_1$-state energy MAE increased by 20-115\% for the SS-$\Delta$, MS-RI, and MS-$\Delta$ models, relative to their respective baselines trained from scratch. Multi-state models with random weight initialization exhibited the poorest performance and highest variance, indicating inherent instability with limited data. The most accurate excited-state potential energy surfaces were achieved by combining multi-state and transfer learning. However, this approach caused a significant degradation in ground-state accuracy, suggesting that separate computational strategies may be optimal for describing the ground and excited states. Notably, 
the performance metrics of our models are competitive with the transferable excited-state machine learning potential X-MACE for the methaniminium cation.\cite{barrett2025transferable}

\begingroup
\squeezetable
\begin{table}[!t]
\centering
\caption{Predictive performance of different models. Shown are the mean absolute errors with standard deviations for energies (meV/atom) and forces (meV/\AA) on the XMCQDPT2 test set for the S$_0$, S$_1$, and S$_2$ states. \label{tab:metrics}}
\begin{ruledtabular}
\begin{tabular}{ldddddd}
\toprule
\multirow{2}{*}{Model} & 
\multicolumn{2}{c}{S$_0$} & 
\multicolumn{2}{c}{S$_1$} & 
\multicolumn{2}{c}{S$_2$} \\
 & \multicolumn{1}{r}{Energy} & \multicolumn{1}{c}{Force} & \multicolumn{1}{r}{Energy} & \multicolumn{1}{c}{Force} & \multicolumn{1}{r}{Energy} & \multicolumn{1}{c}{Force} \\
\cline{2-3} \cline{4-5} \cline{6-7}
SS-RI & 6.3\pm1.6 & 134.1\pm7.3 & 10.8\pm0.7 & 269.5\pm8.3 & 36.2\pm11.4 & 323.6\pm11.3 \\
SS-TL & 4.5\pm0.1 & 94.6\pm4.8 & 8.6\pm0.3 & 201.7\pm4.0 & 28.5\pm4.3 & 258.6\pm3.2 \\
SS-$\Delta$ & 6.2\pm0.6 & 117.1\pm3.8 & 12.8\pm1.4 & 183.6\pm7.1 & 27.6\pm5.8 & 163.1\pm 9.3 \\
MS-RI & 12.6\pm3.7 & 196.8\pm12.5 & 17.2\pm2.1 & 311.8\pm15.5 & 34.4\pm6.2 & 310.6\pm13.4 \\
MS-TL & 9.8\pm0.1 & 106.4\pm0.3 & 6.6\pm0.1 & 188.1\pm0.2 & 20.7\pm0.1 & 222.5\pm0.5 \\
MS-$\Delta$ & 15.6\pm2.5 & 127.0\pm3.6 & 23.2\pm1.7 & 185.4\pm2.3 & 32.6\pm3.8 & 157.7\pm2.2 \\
\bottomrule
\end{tabular}
\end{ruledtabular}
\end{table}
\endgroup

Our subsequent analysis has three key objectives: (1) to rigorously characterize the population dynamics using the SS-TL approach, which demonstrates the best performance across the various electronic states; (2) to evaluate if its superior training metrics yield more reliable dynamics by comparison with an SS-RI baseline; and (3) to compare the resulting average hopping times and product branching ratios across MLIP models.

\subsection{SS-TL model performance on conical intersections}
\begin{figure*}[!h]
    \centering
    \includegraphics[width=0.9\textwidth]{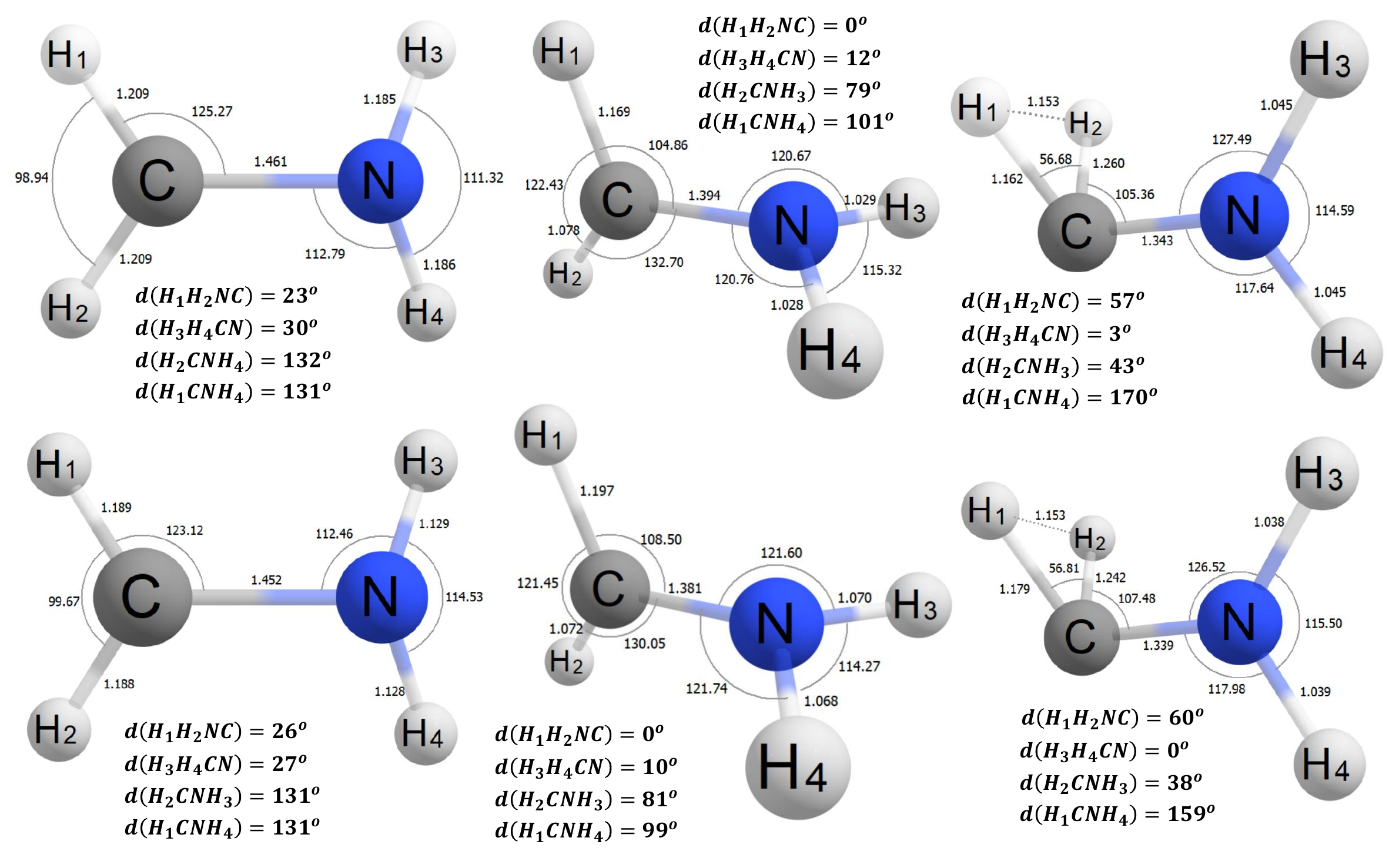}
    \caption{Minimum-energy conical intersections (MECIs) located using an ensemble of the two best-performing SS-TL models (top) and the single best-performing SS-TL model (bottom). Structures shown from left to right are the \ssci{}, \ppci{}, and \spci{} MECIs.} 
    \label{fig:SS-TL-conics}
\end{figure*}

We validated the trained models by applying them to locate the key conical intersections of the \mic{} cation: the S$_2$/S$_1$, \spci{}, and \ppci{} MECIs. Using the Lagrange multiplier method with both an SS-TL ensemble and the single best SS-TL model, we obtained the structures shown in Figure \ref{fig:SS-TL-conics}. Table~\ref{table:conic_vertical_energies} reports the absolute energies and root-mean-square deviations (RMSD) of the optimized geometries, providing a direct comparison against the XMCQDPT2 reference MECI structures.\cite{Bochenkova_JPCL2025} The best SS-TL model achieves high accuracy, predicting energies within 0.2~eV and geometries with RMSD under 0.1~\AA~of the reference. 

\begingroup
\squeezetable
\begin{table}[!h]
\centering
\caption{Absolute energies (a.u.) and root-mean-square deviations (\AA) of the MECIs optimized with the SS-TL models, compared to the XMCQDPT2 reference structures. Standard deviations were calculated from an ensemble of five model predictions.
\label{table:conic_vertical_energies}}
\begin{ruledtabular}
\begin{tabular}{cccccc}
\toprule
\multirow{2}{*}{MECI type} & \multicolumn{2}{c}{Best SS-TL model} & \multicolumn{2}{c}{Ensemble SS-TL} & \multicolumn{1}{c}{XMCQDPT2} \\
& \multicolumn{1}{c}{Energy} & \multicolumn{1}{c}{RMSD} & \multicolumn{1}{c}{Energy} & \multicolumn{1}{c}{RMSD} &\multicolumn{1}{c}{Energy}\\
\hline
 \ppci & -94.5804$\pm$0.0007 & 0.04 & -94.5812$\pm$0.0007 & 0.09 & -94.589587 \\ 
 \spci & -94.5035$\pm$0.0006 & 0.06 & -94.5018$\pm$0.0012 & 0.08 & -94.504066 \\
 \ssci & -94.4308$\pm$0.0008 & 0.07 & -94.4168$\pm$0.0010 & 0.09 & -94.438479 \\
\bottomrule
\end{tabular}
\end{ruledtabular}
\end{table}
\endgroup
The \spci{} MECI is central to a novel nonadiabatic pathway in the methaniminium cation.\cite{Bochenkova_JPCL2025} Figures~S5 and S6 in the Supplementary Material shows potential energy surface scans near this MECI from XMCQDPT2 and the SS-TL model. Importantly, the MLIP reproduces the characteristic double-cone topography of the conical intersection, providing a qualitatively correct representation of this key photochemical feature.

\subsection{Nonadiabatic dynamics and photoproduct branching ratios}

The XMCQDPT2-based ML/LZBL dynamics were initiated from the optically bright S$_2$ ($\pi\pi^*$) state, followed by ultrafast internal conversion to the valence S$_1$ state via the peaked \ssci{} conical intersection. In S$_1$, relaxation proceeds along two competing pathways mediated by the \ppci{} and \spci{} conical intersections. The former is the predominant isomerization channel. The latter, discovered in our previous work~\cite{Bochenkova_JPCL2025}, constitutes a minor dissociation channel leading to direct H$_2$ elimination. In the hot ground state, other dissociation channels also become operational, as substantial excitation energy drives statistical fragmentation in S$_0$ following internal conversion. The initial excitation to S$_2$ specifically promotes CN bond dissociation, as the CN stretching mode is highly active in the Franck-Condon region. Its excitation directly channels the system into the peaked S$_2$/S$_1$ conical intersection, and the resultant nuclear momentum then drives CN bond dissociation upon passage through the $\pi\pi^*$/S$_0$ intersection seam~\cite{barbatti2006ultrafast,barbatti2008nonadiabatic,fabiano2008approximate}.

\begin{figure*}[!b]
    \centering
    \includegraphics[width=1.0\textwidth]{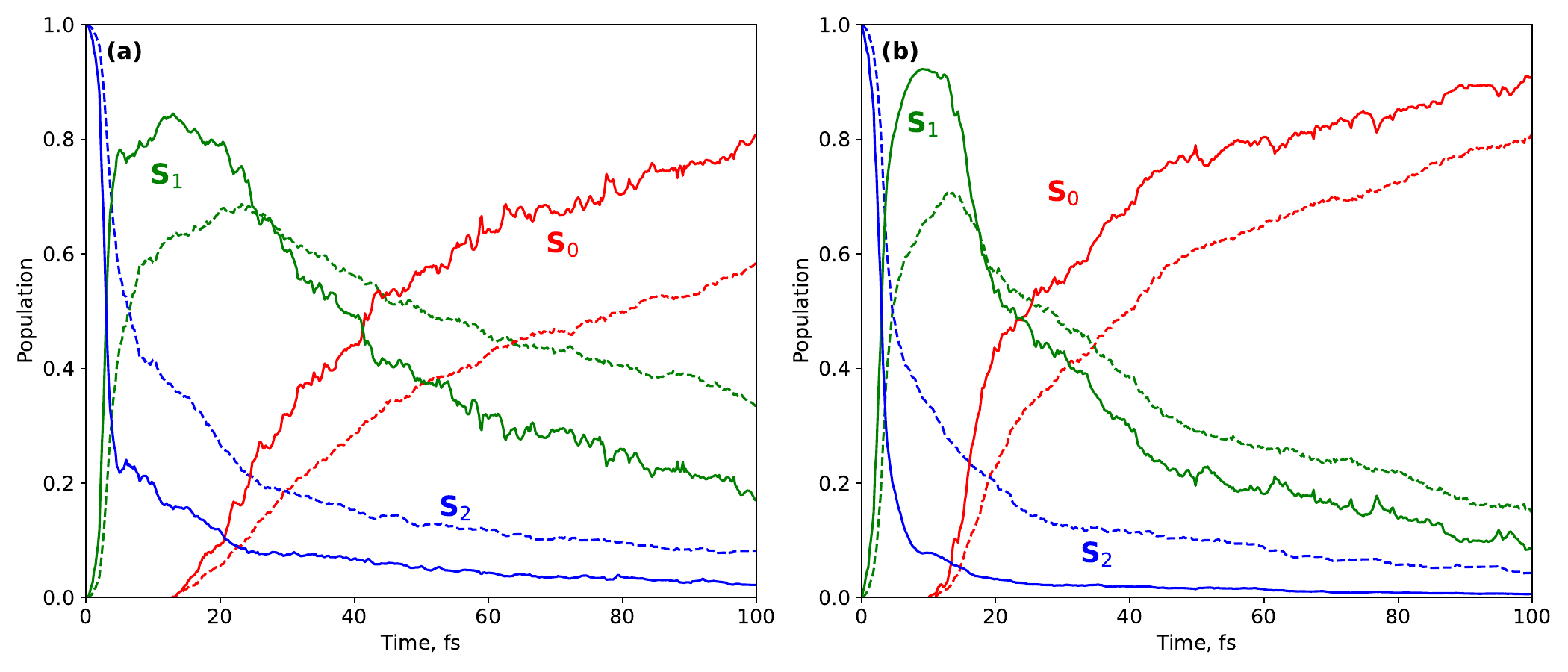}
    \caption{Time evolution of electronic state populations for dynamics initiated in the S$_2$ state, computed using the best SS-RI model (a) and an ensemble of the two best SS-TL models (b). The dashed lines show the average over 600 trajectories. The solid lines represent the average from the simulations corrected for the MLIP uncertainty.} 
    \label{fig:SS-TL-NAMD-S2}
\end{figure*}

Figure~\ref{fig:SS-TL-NAMD-S2} shows the time-dependent state populations for nonadiabatic dynamics following S$_2$ photoexcitation. Using the best-performing SS-TL approach, we obtain uncertainty-corrected average hopping times of 5.0 $\pm$ 0.4~fs for the S$_2$ $\rightarrow$ S$_1$ transition and 28.4 $\pm$ 2.2~fs for the S$_1$ $\rightarrow$ S$_0$ transition (the average time between the S$_2$ $\rightarrow$ S$_1$ hop and the subsequent S$_1$ $\rightarrow$ S$_0$ hop). The S$_2$ $\rightarrow$ S$_1$ internal conversion rate is similar to the on-the-fly CASSCF dynamics (6.0~fs)\cite{Bochenkova_JPCL2025}, but the S$_1$ hopping time is 1.6 times longer than the CASSCF value (18.1~fs)\cite{Bochenkova_JPCL2025}. The application of uncertainty corrections has a pronounced effect: while the uncorrected average hopping times are 14.6 $\pm$ 1.7~fs (S$_2$) and 28.2 $\pm$ 2.9~fs (S$_1$), the corrections notably reduce the S$_2$ lifetime as well as the associated errors, demonstrating their necessity for reliable kinetic modeling. The population dynamics of the superior SS-TL model also differs markedly from the SS-RI baseline (Fig.~\ref{fig:SS-TL-NAMD-S2}). This contrast demonstrates that the predicted dynamics is sensitive to the training metrics, underscoring that the nonadiabatic outcomes are not model-independent. However, the average hopping times show less sensitivity to the choice of the MLIP model. Table~\ref{tab:lifetimes} compares the average hopping times calculated with various MLIP models for dynamics initiated from the S$_2$ state. 
Predictions for the short-lived S$_2$ state are consistent across models, while larger variations are observed for the longer-lived S$_1$ state. Remarkably, the application of the MLIP-uncertainty corrections brings the predictions of the different models into closer agreement, underscoring the value of this diagnostic in improving the reliability and consistency of the dynamics.

\begingroup
\squeezetable
\begin{table}[!t]
\centering
\caption{Uncertainty-corrected and uncorrected average hopping times obtained using various MLIP models for dynamics initiated from the S$_2$ state. \label{tab:lifetimes}}
\begin{ruledtabular}
\begin{tabular}{lcccc}
\toprule
\multirow{2}{*}{Model} & \multicolumn{2}{c}{S$_2$ state } & \multicolumn{2}{c}{S$_1$ state }  \\
 & uncorrected & corrected & uncorrected & corrected  \\
\hline
 SS-RI best model & 11.4$\pm$1.3 & 5.9$\pm$0.5  & 40.5$\pm$2.9 & 37.8$\pm$2.7 \\ 
 SS-RI ensemble   & 12.6$\pm$1.5 & 8.0$\pm$1.2  & 33.8$\pm$2.8  & 32.5$\pm$2.7\\ 
 SS-TL best model & 12.6$\pm$1.5 &  5.0$\pm$0.4 & 29.6$\pm$2.7  & 28.4$\pm$2.1\\ 
 SS-TL ensemble   & 14.6$\pm$1.7 & 5.2$\pm$0.5 & 28.2$\pm$2.9 & 29.0$\pm$2.5 \\ 
 MS-TL best model & 10.6$\pm$1.2 &  -- & 22.4$\pm$2.0  & --\\ 
 MS-TL ensemble   & 10.2$\pm$1.1 &  -- & 21.6$\pm$2.0 &  --\\ 
\bottomrule
\end{tabular}
\end{ruledtabular}
\end{table}
\endgroup

Our dynamics simulations identify seven distinct photodissociation channels:
\begin{enumerate}[leftmargin=*,topsep=0pt,itemsep=-1ex,partopsep=1ex,parsep=1ex]
    \item CN bond cleavage: \mic{} $\rightarrow$ CH$_2^+$ $+$ NH$_2$ 
    \item Concerted H$_2$ / double H-atom elimination from carbon: \mic{} $\rightarrow$ CNH$_2^+$ $+$ H$_2$ / 2H
    \item Double H-atom elimination from nitrogen: \mic{} $\rightarrow$ NCH$_2^+$ $+$ 2H
    \item H-atom elimination from carbon: \mic{} $\rightarrow$ HCNH$_2^+$ $+$ H
    \item H-atom elimination from nitrogen: \mic{} $\rightarrow$ H$_2$CNH$^+$ $+$ H
    \item Double H-atom elimination from C and N: \mic{} $\rightarrow$ HCNH$^+$ $+$ 2H
    \item Non-dissociation: \mic{} remains intact within 100~fs.
\end{enumerate}

\begin{figure*}[!t]
    \centering
    \includegraphics[width=1.0\textwidth]{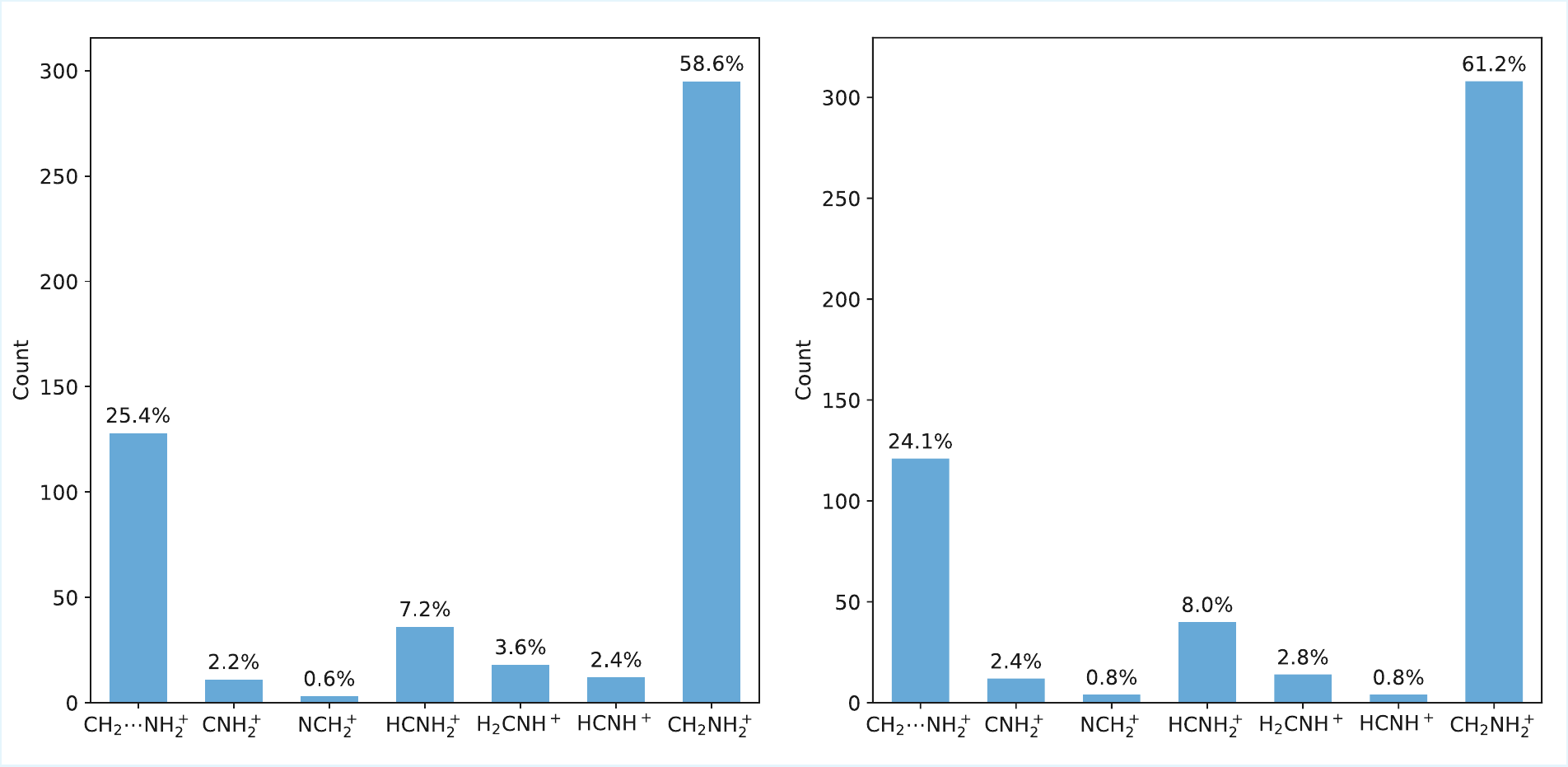}
    \caption{Branching ratios of the photodissociation channels after 100~fs, calculated using an ensemble of the two optimal SS-TL models (left) and the single best SS-TL model (right).} 
    \label{fig:BR}
\end{figure*}

The trajectories were assigned to these groups using the procedure outlined in the Methods section. Figure~\ref{fig:BR} shows branching ratios of the photodissociation channels obtained using the SS-TL models. The predominant dissociation pathway is CN bond cleavage, a direct consequence of its Franck-Condon activation upon S$_0$ $\rightarrow$ S$_2$ photoexcitation. The direct H$_2$ loss channel proceeds via formation of a CNH$_2^+$ carbene intermediate, which then undergoes barrierless isomerization to HCNH$^+$. Within the 100~fs simulation timeframe, this channel accounts for 2\% of the product yield, in agreement with the 2\% branching ratio previously observed at the CASSCF level\cite{Bochenkova_JPCL2025}. Note that unlike other dissociation channels, direct H$_2$ loss is mediated by ultrafast decay through the \spci{} conical intersection. This channel is further characterized by analyzing the population dynamics (see Section~\ref{sec:Model}). 

The product branching ratios show little variation across MLIP models, as they are governed by rapid statistical dissociation on the hot ground-state surface within the first 100~fs, given a large excess of energy (Supplementary Material, Figs.~S7-S8). Among all models, the SS-RI baseline produces the most pronounced deviations, with an approximately 10\% higher yield of intact molecular cations compared to more advanced methods.

\subsection{Wavepacket oscillation model for fitting excited-state population dynamics}
\label{sec:Model}

To extract state-specific lifetimes from first-principles dynamics, we have developed a wavepacket oscillation model for population transfer in three-level systems undergoing sequential nonadiabatic transitions through conical intersections. The population decay is governed by repeated passages through each conical intersection. It moves beyond treating each passage as independent and captures the decreasing probability of survival after multiple attempts. This is fundamentally different from standard kinetics. In this case, the system remembers how many times it has attempted the transition. Each failed hopping attempt increases the probability of successful hopping in subsequent attempts, leading to a power-law decay rather than simple exponential kinetics. The model combines Landau-Zener transition probabilities with wavepacket oscillation dynamics and incorporates ensemble averaging to account for inhomogeneous broadening effects.

We consider a three-level quantum system (S$_2$, S$_1$, and S$_0$) undergoing sequential nonadiabatic transitions. The nonadiabatic cascade is mediated by three conical intersections: first, between S$_2$ and S$_1$, and then by two distinct intersections, labeled \textit{a} and \textit{b}, which provide parallel decay routes from S$_1$ to S$_0$ through the \ppci{} and \spci{} conical intersections, respectively. The population transfer is driven by oscillating wavepackets that sample these intersections.

The time evolution of populations in the three-level system is governed by:
\begin{equation}
P_{S_2}(t) = \left(1 - P_{\text{LZ}}^{(1)}\right)^{N_1(t)}    
\end{equation}
\begin{equation}
P_{S_1}(t) = \alpha\cdot \left(1 - P_{\text{LZ}}^{(a)}\right)^{N_{a}(t)} \cdot \frac{1 - r_a^{N_1(t)}}{1-r_a} + 
(1-\alpha)\cdot \left(1 - P_{\text{LZ}}^{(b)}\right)^{N_{b}(t)} \cdot \frac{1 - r_b^{N_1(t)}}{1-r_b}     
\end{equation}
\begin{equation}
P_{S_0}(t) = 1 - P_{S_2}(t) - P_{S_1}(t), 
\end{equation}
where $P_{S_N}(t)$ is the time-dependent population of the $N$-th electronic state; $P_{\text{LZ}}^{(1)}$, $P_{\text{LZ}}^{(a)}$, and $P_{\text{LZ}}^{(b)}$ are the single-passage transition probabilities; $N_1$(t), $N_{a}$(t) and $N_{b}$(t) -- the effective number of passages through conical intersections \ssci{} ($1$), \ppci{} ($a$), and \spci{} ($b$) by time $t$, respectively; and $\alpha$ is the fraction of the population that goes via channel $a$ in the S$_1$ state. The variable $r_\zeta$ ($\zeta = a, b$) is defined as follows:

\begin{equation}
    r_\zeta= \frac{1-P_{LZ}^{(1)}}{\left(1-P_{LZ}^{(\zeta)}\right)^{T_1/T_\zeta}}
\end{equation}
where $T_1$ and $T_\zeta$ ($\zeta=a, b$) are the oscillation periods around the conical intersections.

The effective number of passages are defined as:
\begin{equation}
N_1(t) = \frac{2(t - t_{\text{ind}}^{(1)})}{T_1} \cdot \Theta(t - t_{\text{ind}}^{(1)}) 
\end{equation}
\begin{equation}
N_{\zeta}(t) = \frac{2(t - t_{\text{ind}}^{(1)} - t_{\text{ind}}^{(\zeta)})}{T_\zeta} \cdot \Theta(t - t_{\text{ind}}^{(1)} - t_{\text{ind}}^{(\zeta)})
\end{equation}
where $t_{\text{ind}}^{(1)}$ and  $t_{\text{ind}}^{(\zeta)}$ ($\zeta=a, b$) are the induction times for each transition, $\Theta$ is the Heaviside step function. The induction times characterize the initial dynamics before significant non-adiabatic transitions occur, while the oscillation periods reflect the curvature of the potential energy surfaces near the conical intersections. 

This approach provides a physically transparent framework for analyzing non-adiabatic dynamics while maintaining computational efficiency through analytical solutions. The model successfully captures the essential features of population transfer in multi-level quantum systems with conical intersections: single-passage probabilities $P_{\text{LZ}}$ and mean times between passages $\tau_{attempt}$. Since a wavepacket, which oscillates with period $T$, passes twice through the intersection region per oscillation cycle, $\tau_{attempt}=T /\ 2$. The probability that the hop occurs precisely on the $N$-th attempt equals to $(1-P_{LZ})^{(N-1)}P_{LZ}$, and the expected value of the number of trials until the first success, including the successful trial, is defined as the mean value of the geometric distribution: 
\begin{equation}
    <N> = \sum_{N=1}^{\infty} {N (1-P_{LZ})^{N-1}P_{LZ}} = \frac{1}{P_{LZ}}
\end{equation}
Therefore, the mean lifetime is defined as:
\begin{equation}
\tau=\tau_{attemp}<N>=\tau_{attemp} /\ P_{LZ}=T /\ 2P_{LZ}. 
\end{equation}
This relation transforms the single-passage quantum mechanical probability $P_{\text{LZ}}$ into a kinetic parameter, the state-specific lifetime $\tau$, that describes the population decay over multiple oscillations.

In the limit of a short oscillation period, $T\rightarrow 0$, and a low single-passage Landau-Zener probability, $P_{\text{LZ}}\rightarrow 0$, the discrete stochastic process converges to a continuous exponential decay. Under these conditions, the number of attempts $N(t)$ within a finite time interval becomes large, while the probability of success per attempt remains small. Mathematically, this is the regime where the expression $(1 - P_{LZ})^{N(t)}$ with $P_{LZ} \ll 1$ and $N(t) \gg 1$ converges to $\exp(-N(t)P_{LZ})$. Applying this limit yields $P(t) \approx \exp[-k_{eff}t]$ with an effective rate constant $k_{\text{eff}} = 2P_{\text{LZ}}/T$. This result establishes the classical rate constant and the corresponding lifetime, defined as its inverse, as derived quantities, rather than empirical parameters. Within the derived multi-passage surviving model, the exact lifetime, defined as the time for the state population to decay by a factor of $1 /\/e$, is given by $\tau_{1/\/e}=-1 /\/2 \cdot T /\/ln(1-P_{LZ})$. Notably, this lifetime matches the average lifetime  $\tau = T/\/2P_{LZ}$, as derived from a geometric distribution of hopping times, only in the limiting case $P_{\text{LZ}}\rightarrow 0$. This quantitative bridge between the quantum mechanical single-event probability and the classical rate constant provides a framework for interpreting ultrafast nonadiabatic dynamics within conventional kinetic language while preserving the essential quantum mechanical character of the individual transitions.

To account for inhomogeneous broadening effects, we perform a statistical average over multiple stochastic realizations of the wavepacket dynamics. Specifically, we generate an ensemble of $\mathcal{N} = 1000$ independent trajectories. For each trajectory $j$, the oscillation periods for the different transitions are independently sampled from normal distributions centered at the mean values $\bar{T}_1$ and $\bar{T}_\zeta$, with relative standard deviations defined by $\sigma_{T_1} = 0.13$ and $\sigma_{T_{\zeta}} = 0.20$.
The final time-dependent population for each state $i$ is obtained by averaging the results from all trajectories:
\begin{equation}
\langle P_i(t) \rangle = \frac{1}{\mathcal{N}} \sum_{j=1}^{\mathcal{N}} P_i^{(j)}(t; T_1^{(j)}, T_a^{(j)}, T_b^{(j)}).
\end{equation}
This numerical averaging procedure is equivalent to integrating over the assumed underlying distribution of periods. The parameters $\sigma_{T_1}$ and $\sigma_{T_{\zeta}}$ thus control the degree of period dispersion within the ensemble, effectively modeling the damping of oscillatory features in the averaged population dynamics.

\begin{figure*}[!t]
    \centering
    \includegraphics[width=0.7\textwidth] {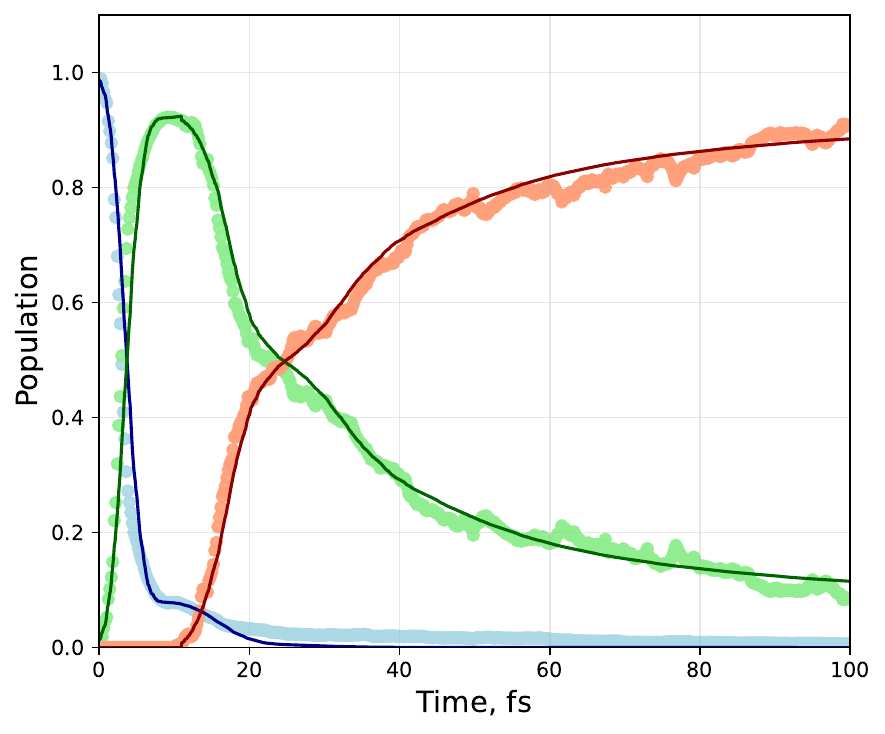}
    \caption{Time evolution of electronic state populations following photoexcitation to S$_2$. The data points (circles) show the average from 600 nonadiabatic trajectories computed with the SS-TL model, corrected for the MLIP uncertainty. The solid lines are the fit obtained from the oscillating wavepacket cascade model, which accounts for branching in decay from S$_1$ to S$_0$.}
    \label{fig:kinetics_fit}
\end{figure*}

\begingroup
\squeezetable
\begin{table}[!t]
\centering
\caption{Fitted kinetic parameters for the multi-passage survival cascade model.
\label{tab:fit_parameters}}
\begin{ruledtabular}
\begin{tabular}{lcccc}
\toprule
\textbf{Process} & $\mathbf{P_{LZ}}$ & $\mathbf{T}$ (\textbf{fs}) &$\boldsymbol{\tau}$ (\textbf{fs}) & $\mathbf{\alpha}$ \\
\hline
$S_2 \rightarrow S_1$             & 0.92 & 25 & $13.6  $  & 1.00 \\
$S_1 \rightarrow S_0$ (Path $a$)  & 0.60 & 30 & $25.0 $ & 0.84 \\
$S_1 \rightarrow S_0$ (Path $b$)  & 0.09 & 40 & $222.2$ & 0.16 \\
\bottomrule
\end{tabular}
\end{ruledtabular}
\end{table}

We employ this model to analyze the population dynamics computed with the SS-TL ensemble of the two best-performing models. The population dynamics is corrected for the MLIP uncertainty. 
Figure~\ref{fig:kinetics_fit} shows the calculated data and the corresponding kinetic fits, and the extracted model parameters are presented in Table~\ref{tab:fit_parameters}.
The oscillating wavepacket model demonstrates high accuracy in reproducing the ultrafast population dynamics. The extracted parameters reveal a three-stage kinetic picture. The initial $S_2 \rightarrow S_1$ relaxation is characterized by the short waiting time $\tau_1 = 13.6$~fs. Due to a high Landau-Zener probability of 0.92, this process nearly completes within one oscillation period, which indicates an extremely efficient, nearly deterministic nonadiabatic transition at the $S_2/S_1$ conical intersection. The subsequent $S_1 \rightarrow S_0$ decay proceeds via two distinct channels with a branching ratio of 84:16. The dominant channel ($a$) is relatively fast with the characteristic time $\tau_{a} = 25.0$~fs and Landau-Zener probability $P^{(b)}_{\text{LZ}}=0.60$, while the minor channel (\textit{b}) is almost an order of magnitude slower with $\tau_{b} = 222.2$~fs and $P^{(b)}_{\text{LZ}}=0.09$. The significant disparity in these characteristic decay times, arising from both the different oscillation periods and transition probabilities, explains the complex character of the $S_1$ decay and the substantial transient population accumulated in this state. The derived model successfully captures the rapid initial decay from $S_2$ and branching in $S_1$, with the period dispersions, $\sigma_{T_{1}}$ and $\sigma_{T_{\zeta}}$, accounting for ensemble effects that smooth out oscillatory features. The accurate fit across all timescales validates the model physical picture, in which the wavepacket oscillates through distinct conical intersections with pathway-specific efficiencies. 

Notably, average hopping times from classical trajectories can deviate from lifetimes extracted from the wavepacket oscillation model for several reasons, including a non-Gaussian distribution of oscillation periods and ensemble variations in hopping probability. In line with these deviations, the fitted lifetime for the fastest S$_2$ $\rightarrow$ S$_1$ transition (13.6~fs) is longer than the average hopping time (5.2 $\pm$ 0.5~fs) extracted from trajectories (Tables~\ref{tab:lifetimes} and~\ref{tab:fit_parameters}). This discrepancy reflects the physical reality that, in the Franck-Condon region close to the \ssci{} conical intersection, attempts are initially concentrated and velocities are high, both of which accelerate hopping. In contrast, the fitted lifetime (25~fs) for the major S$_1$ $\rightarrow$ S$_0$ pathway shows good agreement with trajectory-derived hopping times (29 $\pm$ 2.5~fs), as dynamics occur away from the Franck-Condon region under lower single-passage probabilities. Finally, the lifetime along the minor S$_1$ $\rightarrow$ S$_0$ pathway is the longest; however, because it extends beyond our simulation timescale and is characterized by a small transition  probability, it only marginally increases the overall S$_1$ lifetime derived from the trajectories.

When comparing results with previous works that employ exponential kinetic fits, two critical points must be considered. First, the exponential fits and their derived lifetimes coincide with the average hopping times from a multi-passage survival model only in the limiting case of low transition probability. Second, the specific kinetic scheme used is crucial, as neglecting the minor, delayed decay channel from S$_1$ can artificially extend the fitted S$_1$ lifetime. With these considerations, the previously reported lifetimes, 18.3~fs (S$_2$) and 51.0~fs (S$_1$) obtained from MRCISD/CASSCF(6,4)/aug-cc-pVDZ nonadiabatic dynamics simulations\cite{westermayr2019machine}, are consistent with our results. The lifetimes of $\sim$10~fs (S$_2$) and $\sim$30~fs (S$_1$) are consistent with previous theoretical reports~\cite{Lan_Onthefly21,fabiano2008approximate,barbatti2007fly,barbatti2006ultrafast,FABIANO2008334}.

Importantly, a high-quality fit to the population dynamics is achievable only when the branching in S$_1$ is explicitly included in the kinetic model. This requirement strongly supports the existence of the newly discovered photochemical pathway mediated by a novel \spci{} conical intersection -- a mechanistic insight that would be nearly impossible to deduce from the population evolution alone without the present model. The kinetic fits yield channel-specific lifetimes, for the first time revealing a long timescale ($\sim$200~fs) for the minor pathway. Despite its substantial branching ratio of 16\%, this pathway accounts for only a 2\% photoproduct yield within the first 100~fs.

\section{Conclusions}
In this work, we have developed a transfer-learning protocol to construct a series of machine-learning interatomic potentials that achieve the accuracy of extended multi-configuration quasi-degenerate perturbation theory (XMCQDPT2/ SA(3)-CASSCF(12,12)/ aug-cc-pVDZ). This enabled, for the first time, high-level on-the-fly nonadiabatic dynamics simulations of the methaniminium cation initiated in the S$_2$ state using a large active space. These comprehensive simulations map the system complete photodissociation landscape, capturing the competing decay channels, including the major photoisomerization and minor direct H$_2$-loss pathways. To bridge microscopic nonadiabatic events with ensemble kinetics, we introduced a novel wavepacket oscillation model. This mechanistically transparent, power-law kinetics framework extracts state-specific lifetimes directly from first-principles population dynamics, moving beyond empirical kinetic fitting. It establishes a quantitative link between fundamental quantum transition probabilities and classical rate constants, preserving the quantum character of conical intersection passages within an interpretable kinetic description.

Our analysis reveals that while nonadiabatic outcomes are sensitive to the MLIP training strategy, with the advanced transfer-learning model yielding dynamics distinct from the baseline, key observables such as average hopping times and product branching ratios show notable consistency across models. Applying MLIP-uncertainty corrections from an ensemble of models brings the predictions of different training approaches into closer agreement, validating this metric as an important diagnostic for simulation reliability. Furthermore, high-quality kinetic fits of the population dynamics in the three-level \mic{} system requires explicit inclusion of the S$_1$ branching, providing independent validation for the newly discovered photochemical pathway mediated by a novel \spci{} conical intersection\cite{Bochenkova_JPCL2025}. The kinetic fits resolve a channel-specific lifetime of $\sim$200~fs for this minor pathway, resulting from its low transition probability of 9\%. In contrast, the major pathway exhibits a far shorter S$_1$ lifetime of 25~fs, corresponding to a relatively high transition probability of 60\%. The minor channel accounts for 16\% of the total trajectories.

Collectively, this work delivers three key advances: a robust framework for constructing high-fidelity MLIPs for excited-state dynamics, a first-principles mapping of the complete photodissociation mechanism of a fundamental model system, and a new power-law kinetics theory that directly connects quantum nonadiabatic transitions to interpretable, predictive lifetime models. These contributions establish a generalizable pathway for simulating and analyzing ultrafast photochemical processes with both high accuracy and mechanistic clarity.\\

\newpage
\noindent
{\bf SUPPLEMENTARY MATERIAL}
\noindent \\
The supplementary material provides additional figures: a) predictive performance of various MLIP models; b) time-dependence of the characteristic distance parameters along trajectories; c) comparison of XMCQDPT2 and MLIP potential energy surfaces in the vicinity of the \spci{} conical intersection; d) branching ratios of the photodissociation channels obtained using various MLIP models.\\

\noindent
{\bf ACKNOWLEDGMENTS}
\noindent \\
This work was supported by the Russian Science Foundation (Grant No. 24-43-00041). The calculations were carried out using the equipment of the shared research facilities of HPC computing resources at Lomonosov Moscow State University as well as the local resources (RSC Tornado) provided through the Lomonosov Moscow State University Program of Development.\\

\noindent
{\bf DATA AVAILABILITY}
\noindent \\
The data that support the findings of this study are available from the corresponding author upon reasonable request.
\\

\noindent
{\bf CONFLICT OF INTEREST}
\noindent \\
The authors have no conflicts to disclose.
\\

\noindent
{\bf AUTHOR CONTRIBUTIONS}
\noindent \\
Ivan V. Dudakov: Data curation (equal); Formal analysis (equal); Investigation (equal); Software (equal); Validation (equal); Visualization (equal); Writing – original draft (equal); Writing – review \& editing (supporting); Pavel M. Radzikovitsky: Data curation (equal); Formal analysis (equal); Investigation (equal); Software (equal); Validation (equal); Visualization (equal); Writing – review \& editing (supporting); Dmitry S. Popov: Data curation (equal); Formal analysis (equal); Investigation (equal); Writing – review \& editing (supporting); Denis A. Firsov: Investigation (equal), Visualization (equal); Writing – review \& editing (supporting); Vadim V. Korolev: Conceptualization (equal); Investigation (equal); Methodology (equal); Software (equal); Validation (equal); Resources (equal); Writing – original draft (equal); Writing – review \& editing (supporting);  Daniil N. Chistikov: Conceptualization (equal); Data curation (equal); Formal analysis (equal); Investigation (equal); Methodology (equal); Software (equal); Validation (equal); Visualization (equal); Resources (equal); Writing – review \& editing (supporting); Vladimir V. Bochenkov: Conceptualization (equal); Formal analysis (equal); Investigation (equal); Methodology (equal); Software (equal); Validation (equal); Visualization (equal); Writing – original draft (equal); Writing – review \& editing (supporting); Anastasia V. Bochenkova: Conceptualization (lead); Formal analysis (equal); Funding acquisition (lead); Investigation (equal); Methodology (equal); Project administration (lead); Resources (lead); Supervision (lead); Validation (supporting); Writing – original draft (equal); Writing – review \& editing (lead).


\bibliography{bibliography}

\end{document}